\documentclass[twoside,fleqn]{article}
\usepackage{espcrc2}
\usepackage[dvips]{epsfig}

\title{
  \vskip-2.0ex\hbox to 6.25in {{\normalsize \hfil KEK-CP-093}}
  $B$ decays on the lattice
  }

\author{
  Shoji Hashimoto\address{Computing Research Center, 
    High Energy Accelerator Research Organization (KEK), 
    Tsukuba 305-0801, Japan}
  }

\begin{document}

\begin{abstract}
  I review recent developments in lattice calculations of
  $B$ decay matrix elements and other related quantities. 
\end{abstract} 

\maketitle

\section{Introduction}
\label{sec:Introduction}

$B$ decay phenomenology is a rapidly growing area of
particle physics, as we expect that the precise measurements
of various $B$ meson decays provide rich information of the
Standard Model and the physics beyond it. 
Following the successful CLEO, LEP and CDF experiments, the
next generation $B$ factories (BaBar and Belle) have just
started operation, and other experimental projects will
follow. 

Lattice QCD may contribute to this program by
calculating the $B$ meson decay matrix elements starting
from the first principles.
Its application is extending from the decay constant $f_B$
and bag parameter $B_B$ to various semileptonic decay form
factors.
Results are being checked by several groups using different
lattice formulations of heavy quark, and systematic errors
are carefully estimated.
In addition, studies of new applications have been started,
such as the zero recoil form factor of 
$B\rightarrow D^{(*)} l\nu$, the width difference of $B_s$
meson, and the lifetime ratio of $b$ hadrons.
In this talk I review recent developments in such efforts. 

This talk is organized as follows.
Before discussing the matrix elements calculation, I will
summarize recent progress in the lattice determination of
$b$ quark mass in the next section. 
Semileptonic decays are discussed in Section
\ref{sec:Semileptonic_decays}. 
These are divided into two parts: heavy-to-heavy transitions 
($B\rightarrow D^{*}l\nu$, Section \ref{sec:B->Dlnu}) and 
heavy-to-light decay ($B\rightarrow \pi l\nu$,
Section \ref{sec:B->pilnu}). 
I then describe the $B-\bar{B}$ mixing, where $f_B$ and
$B_B$ are relevant. 
Recent unquenched calculations of $f_B$ is summarized in
Section \ref{sec:f_B}, and some updates on $B_B$ is
discussed in Section \ref{sec:B_B}.
Finally I will discuss new applications of lattice QCD,
which include the calculation of $B_s$ width difference
(Section \ref{sec:B_s_width_difference}) and $b$-hadron
lifetime ratios (Section \ref{sec:Lifeteime_ratios}).
Those calculations involve similar matrix elements as in
$B_B$.
A summary of the status as of 1998 was given by T.~Draper 
at the last lattice conference \cite{Draper_lat98}.

\section{The $b$ quark mass}
\label{sec:b_quark_mass}

Although the $b$ quark mass is not a matrix element related
to any exclusive decay of $B$ meson, it is necessary in the
calculation of many inclusive decay rates.
It has become clear, however, that there is an ambiguity
in the definition of pole mass of heavy quark itself
(renormalon ambiguity), and its determination cannot be
better than of order $\Lambda_{QCD}$ \cite{Beneke_98}.
In order to avoid the ambiguity one has to use a short
distance mass definition, such as $\overline{MS}$ quark mass
$\overline{m}_b(\overline{m}_b)$. 

In the lattice QCD calculation, important progress was made
by Martinelli and Sachrajda \cite{Martinelli_Sachrajda_98},
when they obtained the two-loop coefficient in the
perturbative matching between continuum QCD and lattice HQET
(static approximation) 
\begin{eqnarray}
  \label{eq:b-quark_mass}
  \lefteqn{\overline{m}_b(\overline{m}_b) = \Delta \times 
    \left[ 1 + \frac{1}{\Delta a}
            \{2.1173\alpha_s(\overline{m_b}) \right.} \nonumber\\
  & & \biggl. 
      +(3.707\ln(\overline{m}_b a)-1.306)\alpha_s^2(\overline{m}_b) \}  
      \biggr] \nonumber\\
  & & \times 
      \left[ 1-\frac{4}{3}\frac{\alpha_s(\overline{m}_b)}{\pi}
        -11.66\left(\frac{\alpha_s(\overline{m}_b)}{\pi}\right)^2
      \right],
\end{eqnarray}
where $\Delta=M_B-\mathcal{E}$ is obtained with the
``binding energy'' $\mathcal{E}$ measured in lattice
simulations. 
In this expression the perturbative expansion in the first
parenthesis matches the energy shift in the lattice HQET to 
the heavy quark pole mass.
A power divergence exists in the loop integral that cancels
with power divergent behavior of nonperturbatively
calculated $\Delta$. 
The last line connects the pole mass to the $\overline{MS}$
mass.
The renormalon ambiguity exists in both of these two
expansions, while it cancels between them.
With the perturbative expansion to two-loop, these
cancellations are expected to become more precise, making
the result more stable.

\begin{figure}[tb]
  \begin{center}
%    \vspace*{-0mm}
    \epsfxsize=74mm \epsfbox{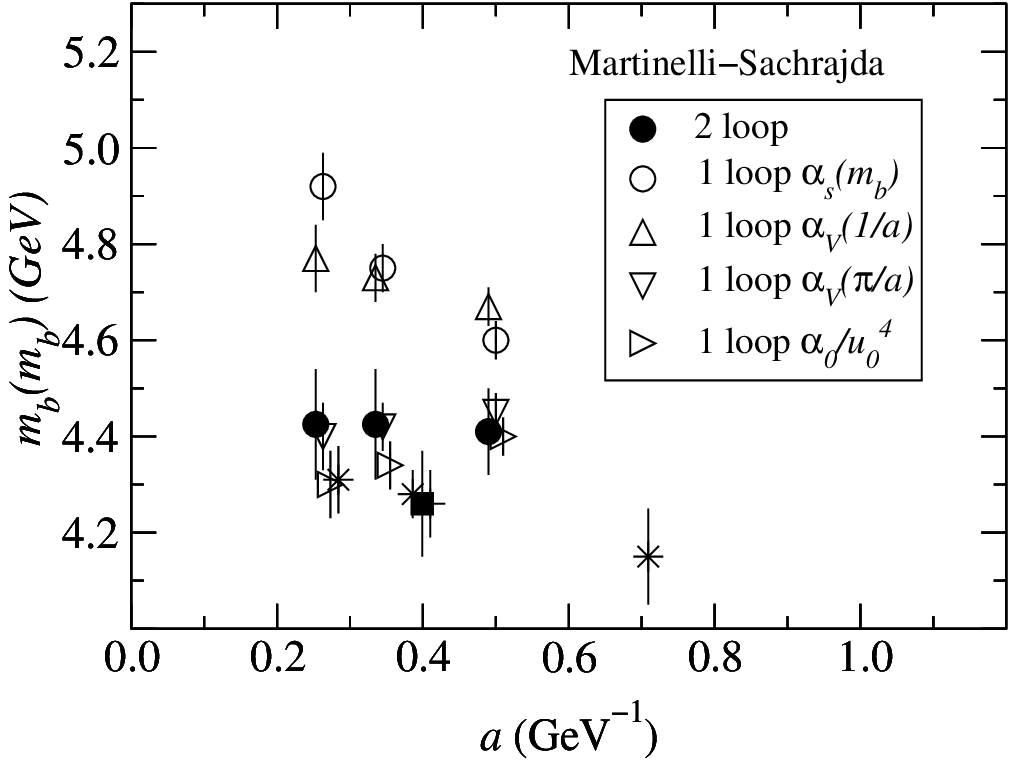}
    \vspace*{-10mm}
    \caption{Lattice calculation of the $b$ quark mass.
      Martinelli-Sachrajda's \cite{Martinelli_Sachrajda_98}
      two-loop results (filled circles) are plotted together
      with the corresponding one-loop matching results with
      various definitions of coupling constant (open
      symbols). 
      APE's HQET result \cite{Gimenez_lat99} with $N_F=2$
      (square), and NRQCD result \cite{NRQCD_lat98} with
      $N_F=0$ (star) and with $N_F=2$ (plus) are also
      shown. 
      }
%    \vspace*{-5mm}
    \label{fig:b-quark_mass}
  \end{center}
\end{figure}

Numerical results with the two-loop matching are very 
stable among three $\beta$ values (6.0, 6.2 and 6.4) as
shown in Figure \ref{fig:b-quark_mass}, where corresponding
one-loop results are also plotted with various definitions of
the coupling constant.
This result suggests that the cancellation of the power
divergence and the renormalon ambiguity is working fairly
well at two-loop level.

A dynamical ($N_F=2$) result along this line has been
reported at this conference by the APE collaboration 
(V. Gim\'enez \cite{Gimenez_lat99}), which is shown by a
square in Figure \ref{fig:b-quark_mass}. 
Their result does not show significant shift from the
quenched results.

The NRQCD collaboration presented a $b$ quark mass calculation 
using NRQCD action at Lattice 98 \cite{NRQCD_lat98}.
They use a one-loop matching between the kinetic mass in NRQCD 
and the $\overline{MS}$ mass, where no power divergence
appears in the perturbative calculation and the one-loop
coefficient is not large.
Their results with (plus) and without (stars) two-flavour
dynamical fermions are consistent with the HQET
calculations. 

The best estimate from the lattice calculations
$\overline{m}_b(\overline{m}_b)$ = 4.26$\pm$0.11 GeV, which I
take from the two dynamical results, is in good agreement
with recent continuum calculations
4.25(8) \cite{Beneke_Signer_99},
4.20(6) \cite{Hoang_99},
4.20(10) \cite{Melnikov_Yelkhovsky_99}
on the $\Upsilon$ resonances, and with
DELPHI's measurement 3.91(67) GeV at $Z$ pole
\cite{DELPHI_98}.

\section{Semileptonic decays}
\label{sec:Semileptonic_decays}

The lattice calculation of $B$ meson semileptonic decay form
factors may be used to determine $|V_{cb}|$ and $|V_{ub}|$
with corresponding experimental results of exclusive decays 
$B\rightarrow D^{(*)} l\nu$ and 
$B\rightarrow \pi(\rho) l\nu$.

\subsection{$B\rightarrow D^{(*)} l\nu$}
\label{sec:B->Dlnu}

\subsubsection{Zero recoil form factor}
One of the most promising methods to determine $|V_{cb}|$ is
to measure 
$|V_{cb}|^2\mathcal{F}_{B\rightarrow D^{(*)}}^2(w)$ 
from the differential decay rate of 
$B\rightarrow D^{(*)} l\nu$, and extrapolate it to the zero 
recoil limit of daughter meson $w\rightarrow 1$, at which
the form factor $\mathcal{F}_{B\rightarrow D^{(*)}}(1)$ is
normalized to unity in the heavy quark mass limit.
Theoretical calculations of the power correction to the
heavy quark limit have been made using QCD sum rule, but
their uncertainty is still quite large ($\sim$ 9\% for 
$\mathcal{F}_{B\rightarrow D^{*}}(1)$) \cite{Bigi_99}.
Lattice calculation could be an important alternative, if it 
provides determination better than $O$(5\%).

Recently, the Fermilab group has proposed a method to
calculate the deviation from the heavy quark limit by
actually measuring the heavy quark mass dependence of the
form factors on the lattice
\cite{Hashimoto_et_al_99,Simone_et_al_lat99}.
Their key observation is that most of statistical and
systematic errors cancel in a ratio of matrix elements
\cite{Hashimoto_et_al_99}
\begin{equation}
  \label{eq:h_+_ratio}
  |h_+^{B\rightarrow D}(1)|^2 =
  \frac{\langle D|V_0^{cb}|B\rangle 
        \langle B|V_0^{bc}|D\rangle}{
        \langle D|V_0^{cc}|D\rangle
        \langle B|V_0^{bb}|B\rangle},
\end{equation}
where $h_+^{B\rightarrow D}(1)$ denotes a zero recoil form
factor of $B\rightarrow Dl\nu$ decay through a vector
current $V_{\mu}^{cb}$.
The statistical error is less than one per cent for typical 
values of $m_b$ and $m_c$, and thus it is possible to study
its mass dependence.
The renormalization factor to match the heavy-heavy current 
on the lattice to its continuum counter part is
perturbatively calculated, depending on the heavy quark mass, 
and found to be small for the above ratio
\cite{Kronfeld_Hashimoto_lat98}. 

They fit their data for $h_+(1)$ as well as for 
$h_1(1)$ and $h_{A_1}(1)$, which are form factors of
$B^*\rightarrow D^*$ and $B\rightarrow D^*$ modes obtained
through similar ratios, with the form of $1/m_Q$ expansion
predicted by the heavy quark symmetry
\cite{Falk_Neubert_93,Mannel_94} 
\begin{eqnarray}
  \label{eq:1/m_expansion}
  h_+(1) 
  &\hspace{-0.5em}=\hspace{-0.5em}&
  1 - l_P
  \left(\frac{1}{2m_c}-\frac{1}{2m_b}\right)^2 + O(1/m_Q^3),
  \nonumber\\
  h_1(1) 
  &\hspace{-0.5em}=\hspace{-0.5em}&
  1 - l_V
  \left(\frac{1}{2m_c}-\frac{1}{2m_b}\right)^2 + O(1/m_Q^3),
  \nonumber\\
  h_{A_1}(1) 
  &\hspace{-0.5em}=\hspace{-0.5em}&
  1 -
  \left(\frac{1}{2m_c}-\frac{1}{2m_b}\right)
  \left(\frac{l_V}{2m_c}-\frac{l_P}{2m_b}\right) \nonumber\\
  & & + \frac{\Delta}{4m_c m_b} + O(1/m_Q^3).
\end{eqnarray}
The parameters $l_P$, $l_V$ and $\Delta$ of $O(1/m_Q^2)$
terms are determined with the fit.
Their preliminary result for $B\rightarrow D^* l\nu$ form
factor obtained at $\beta$=5.7 is 
$\mathcal{F}_{B\rightarrow D^*}(1)$ = 
0.935(22)($^{+8}_{-11}$)(8)(20),
where errors represent statistical, mass determination,
perturbative and unknown $O(1/m_Q^3)$ in the given order
\cite{Simone_et_al_lat99}.
The systematic error associated with the Fermilab formalism of 
heavy quark \cite{El-Khadra_Kronfeld_Mackenzie_97} is a
subtle issue, especially because the $1/m_Q^2$ terms in the
effective Hamiltonian are not correctly tuned with the use
of the clover action, while the correction to be measured is
of $O(1/m_Q^2)$.
Nevertheless, using the $1/m_Q$ expansion, it is generally
shown that only $O(1/m_Q)$ terms in the action and currents
contribute to the ratios calculated above, and thus the
systematic error is well under control 
\cite{Kronfeld_lat99}. 
The result is consistent with the recent QCD sum rule
calculation 0.89$\pm$0.08 \cite{Bigi_99}, and the estimated
error is already slightly smaller, demonstrating a
possibility to improve the determination of $|V_{cb}|$ using 
the lattice calculation in near future.

\subsubsection{Shape of the form factor}
In recent high statistics experiments, the extrapolation
to the zero recoil limit without theoretical
constraints seems good enough \cite{DELPHI_Vcb_99}.
It is, however, informative to compare the shape of form
factors with theoretical predictions in order to check the 
reliability of the extrapolation, and also to check the
theoretical methods. 
Two new calculations of heavy-to-heavy form factors
(Isgur-Wise function) have been presented by UKQCD
\cite{UKQCD_Douglas_lat99} and by Hein \textit{et al.} 
\cite{Hein_et_al_lat99} at this conference.

UKQCD used the non-perturbatively $O(a)$-improved action at
$\beta$=6.0 and 6.2, and calculated $D\rightarrow D$ form
factor $h_+$ \cite{UKQCD_Douglas_lat99}.
They found no significant dependence on the heavy quark
mass, and their preliminary result for the slope parameter
is $\rho^2_{u,d}$ = 
1.10($^{+27}_{-13}$)($^{+7}_{-4}$) and 
1.12($^{+29}_{-15}$)($^{+6}_{-5}$) at $\beta$=6.0 and 6.2,
respectively. 
The precision is the best among previous results
\cite{Flynn_Sachrajda_97}, and the scaling with two lattice
spacings is remarkable.

Hein \textit{et al.} presented a very preliminary
study with a NRQCD action at $\beta$=5.7
\cite{Hein_et_al_lat99}.
Form factor $h_+$ is obtained for $B\rightarrow D$ and also
for $B\rightarrow D'$ mode, where $D'$ denotes a radially
excited state of $D$ meson.
Perturbative matching factors recently calculated by Boyle
\cite{Boyle_lat99} are incorporated for heavy-heavy
(axial-)vector currents.

\subsection{$B\rightarrow \pi l\nu$}
\label{sec:B->pilnu}

Exclusive semileptonic decays $B\rightarrow \pi(\rho)l\nu$
could be used for the determination of $|V_{ub}|$,
provided that corresponding form factors are theoretically
calculated. 
Unfortunately, in the lattice simulations, it is difficult
to put large recoil momentum on the daughter meson, so that
momentum transfer squared $q^2$ is restricted in the large
$q^2$ region (small recoil momentum).
The lattice calculations can be, however, still useful, if
statistics in the experiment is precise enough to measure
the partial decay rate for the large $q^2$ region, and such
work has already been done by CLEO for 
$B\rightarrow\rho l\nu$ \cite{CLEO_99}.

There are two form factors $f^+(q^2)$ and $f^0(q^2)$
involved in $B\rightarrow\pi l\nu$.
$f^0(q^2)$ has a negligible contribution to the physical decay
rate, as it is proportional to the lepton mass.
It is, however, interesting to study $f^0(q^2)$, since
the soft pion theorem relates $f^0(q^2_{max})$ to
$f_B$, which may be used for a test of the lattice
calculations. 
On the other hand, the calculation of $f^+(q^2)$ is of
practical importance, as it is directly related to the
physical decay rate and may be used for a precise
determination of $|V_{ub}|$ of $O(10\%)$.

\subsubsection{$f^0(q^2)$}
Based on the LSZ reduction formula and current algebra, 
the soft pion theorem predicts 
\begin{equation}
  \label{eq:soft_pion}
  f^0(q^2_{max}) = \frac{f_B}{f_{\pi}}
\end{equation}
in the limit $m_{\pi}\rightarrow 0$ and 
$p_{\pi}\rightarrow 0$, thus $q^2_{max}=m_B^2$.

\begin{figure}[tb]
  \begin{center}
%    \vspace*{-0mm}
    \epsfxsize=74mm \epsfbox{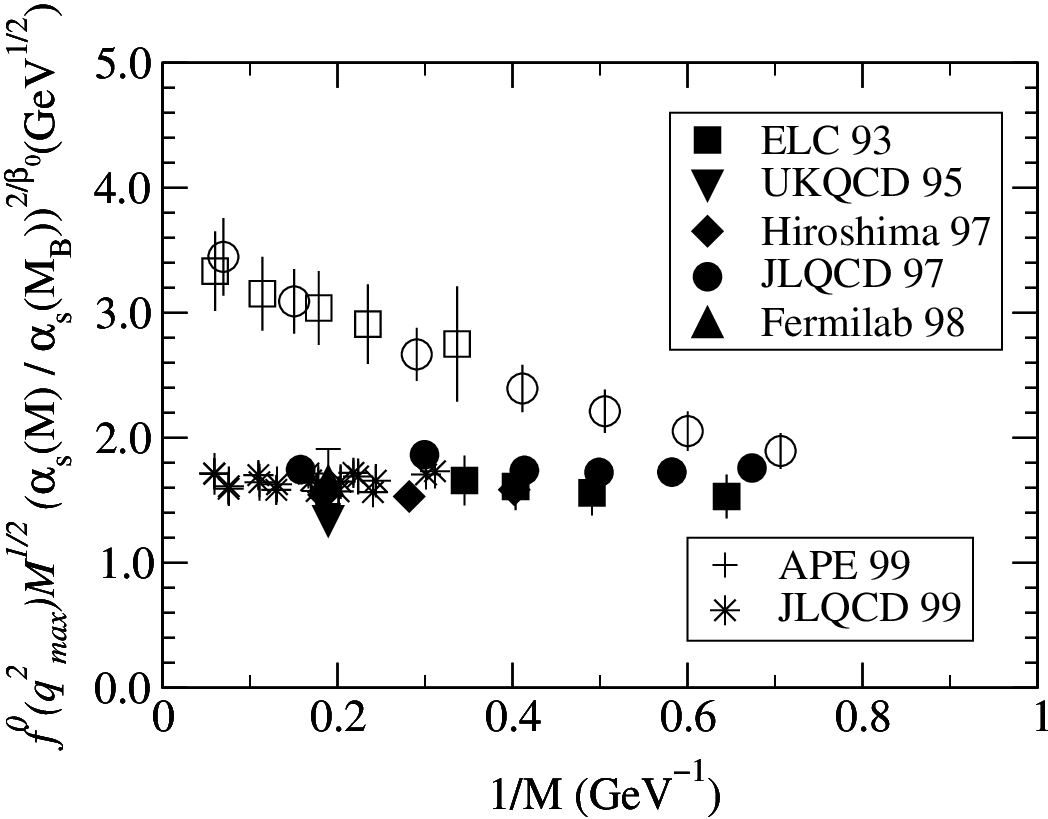}
    \vspace*{-10mm}
    \caption{Check of the soft pion relation.
      Data for $f^0(q^2_{max})$ are from
      ELC \cite{ELC_B2pi_93}, 
      UKQCD \cite{UKQCD_B2pi_95},
      Hiroshima \cite{Hiroshima_B2pi_97}, 
      JLQCD \cite{JLQCD_B2pi_lat97},
      Fermilab \cite{Fermilab_B2pi_lat98}.
      Also plotted are new results reported at this
      conference by
      APE \cite{APE_B2pi_lat99} and 
      JLQCD \cite{JLQCD_B2pi_lat99}. 
      Light quark mass is fixed at the strange quark mass
      for some of the above data, since there is no
      significant light quark mass dependence found so far.
      Two representative results are taken for $f_B/f_{\pi}$ 
      from JLQCD with clover action (open circles)
      \cite{JLQCD_fB_97} and with NRQCD action (open
      squares) \cite{JLQCD_fB_99}.
      }
%    \vspace*{-5mm}
    \label{fig:soft_pion}
  \end{center}
\end{figure}

It was pointed out by Onogi at Lattice 97 
that the lattice results for $f^0(q^2_{max})$ and $f_B$ do
not seem to satisfy this relation \cite{Onogi_lat97}.
The inconsistency has become even clearer with new
data as shown in Figure \ref{fig:soft_pion}, in which I plot
currently available results for $f^0(q^2_{max})\sqrt{M}$
(with renormalization group correction), comparing them with 
representative data for $(f_B/f_{\pi})\sqrt{M}$.
The heavy quark scaling law predicts both quantities should
be constant, up to $1/M$ corrections.

Possible cures for this problem have been proposed at this
conference: nonperturbative renormalization and a method of
chiral and $q^2$ extrapolation, which I discuss in the
following.

The JLQCD collaboration has calculated a ratio of
renormalization factor $(Z_A/Z_V)^{HL}$ nonperturbatively, 
using the chiral Ward identity \cite{JLQCD_B2pi_lat99}
\begin{eqnarray}
  \label{eq:chiral_Ward_identity}
  \lefteqn{
    Z_A Z_V^{HL} \int d^4y \langle
    (\partial_{\mu}A_{\mu}-2m_q P)(y) V_0^{HL}(x)
    \cal{O} \rangle 
    } \nonumber \\
    & = & - Z_A^{HL} \langle A_0^{HL}(x) \cal{O} \rangle,
\end{eqnarray}
where $A_{\mu}$ and $P$ denote the light-light axial-current
and pseudoscalar density defined on the lattice, and $Z_A$
represents the renormalization factor for $A_{\mu}$
available nonperturbatively \cite{ALPHA_96}.
$V_0^{HL}$ and $A_0^{HL}$ are temporal component of
heavy-light vector and axial-vector currents, which appear
in the definition of $f^0(q^2_{max})$ and of $f_B$
respectively. 
Their simulation method follows that of Maiani and
Martinelli \cite{Maiani_Martinelli_86}, while they use
$O(a)$-improved action and current for light quark and
static heavy quark action.
A preliminary result for the static-light currents at
$\beta$=6.0, $(Z_A/Z_V)^{HL}$=0.72(1), is significantly
smaller than the one-loop value 0.86(4).
This result indicates that the one-loop calculation of the
renormalization factor of the static-light current may
contain a large systematic error ($\sim$ 20\%), and suggests
larger $f^0(q^2_{max})$ and/or smaller $f_B$ in the static
limit, which gives right direction to satisfy the soft pion
relation, but magnitude is not enough.
%I also note that the static-light current is not improved
%with higher dimensional operator, which is known to give
%a large contribution in the calculation of $f_B$
%\cite{Morningstar_Shigemitsu_98}. 
An ongoing study to calculate $Z_A^{HL}$ nonperturbatively
using the Scr\"odinger functional method has also been
presented at this conference by Kurth and Sommer
\cite{Kurth_lat99}.

UKQCD has pointed out that a term proportional to $m_{\pi}$
is necessary in the chiral extrapolation as well as the
usual $m_{\pi}^2$ term \cite{UKQCD_B2pi_lat99}. 
That dependence appears solely from the $m_{\pi}$ dependence
of $q^2_{max}$: 
$q^2_{max} = (m_B-m_{\pi})^2 \simeq m_B^2-2m_B m_{\pi}$.
As a result the chiral extrapolation could be subtle, since
$m_{\pi}$ and $m_{\pi}^2$ behave similarly in the region
where simulations are made.
In order to remove $m_{\pi}$ term they interpolate
$f^0(q^2)$ to several fixed $q^2$ values for each of the
active and spectator light quark masses. 
Then, the chiral extrapolation can be done at each fixed
$q^2$ with the $m_{\pi}^2$ term only.
The range of available $q^2$ for all light quark mass is
substantially lower than the physical $q^2_{max}$, so they employ 
a pole dominance model
\footnote{Strictly speaking they combine $f^+$ and $f^0$
  when they fit with dipole-pole ansatz \cite{UKQCD_B2pi_95}
  or with Becirevic-Kaidalov model
  \cite{Becirevic_Kaidalov_99}. 
  The functional form of $f^0(q^2)$ are pole-type in both
  models.} 
to obtain the physical $f^0(q^2_{max})$ from data points.
This extrapolation to the physical $q^2_{max}$ makes
$f^0(q^2_{max})$ significantly high
1.3($^{+3}_{-2}$)($^{+3}_{-2}$), which is consistent with
$f_B/f_{\pi}$=1.35($^{+5}_{-5}$)($^{+11}_{-7}$).

Although it is a nice observation, the method is, to some
extent, model dependent.
Therefore, a cross check seems necessary with the direct 
extrapolation method including $m_{\pi}$ and $m_{\pi}^2$
terms, which requires high statistics data at several
(active and spectator) light quark masses.

MILC collaboration have started such a study using fatlink clover 
quark action for both heavy and light quarks on their
dynamical quark configurations
\cite{MILC_semileptonic_lat99}. 
They presented a very preliminary result that soft pion
theorem is satisfied when extrapolated including $m_{\pi}$
term albeit with a large error.

\subsubsection{$f^+(q^2)$}
Model independent calculation of $f^+(q^2)$ has great
phenomenological importance, since the differential decay
rate $d\Gamma(B\rightarrow\pi l\nu)/dq^2$ is proportional to 
$|V_{ub}|^2 |f^+(q^2)|^2$ and its measurement may be used
for the determination of $|V_{ub}|$.

Near $q^2_{max}=(m_B-m_{\pi})^2$, where lattice calculation
is most powerful, the pole dominance picture should give a
good approximation, as the corresponding pole of $B^*$ meson 
is very close.
Using the $B^*B\pi$ coupling $g_{B^*B\pi}$ and $B^*$ meson
decay constant $f_{B^*}$, defined by
\begin{eqnarray}
  \label{eq:B*B_pi_definition}
  \langle B(p)\pi(k)|B^*(p')\rangle 
  & = & g_{B^*B\pi} (k\cdot\epsilon), \\
  \langle 0|V_{\mu}|B^*\rangle 
  & = & i f_{B^*}m_{B^*} \epsilon,
\end{eqnarray}
the form factor $f^+(q^2)$ is
given by
\begin{equation}
  \label{eq:pole_dominance}
  f^+(q^2) = 
  \frac{\frac{1}{2} g_{B^*B\pi} f_{B^*} m_{B^*}}{
    m_{B^*}^2-q^2}
\end{equation}
up to the contribution of higher excited states, which is
relatively small for large $q^2$.

\begin{figure}[tb]
  \begin{center}
%    \vspace*{-0mm}
    \epsfxsize=74mm \epsfbox{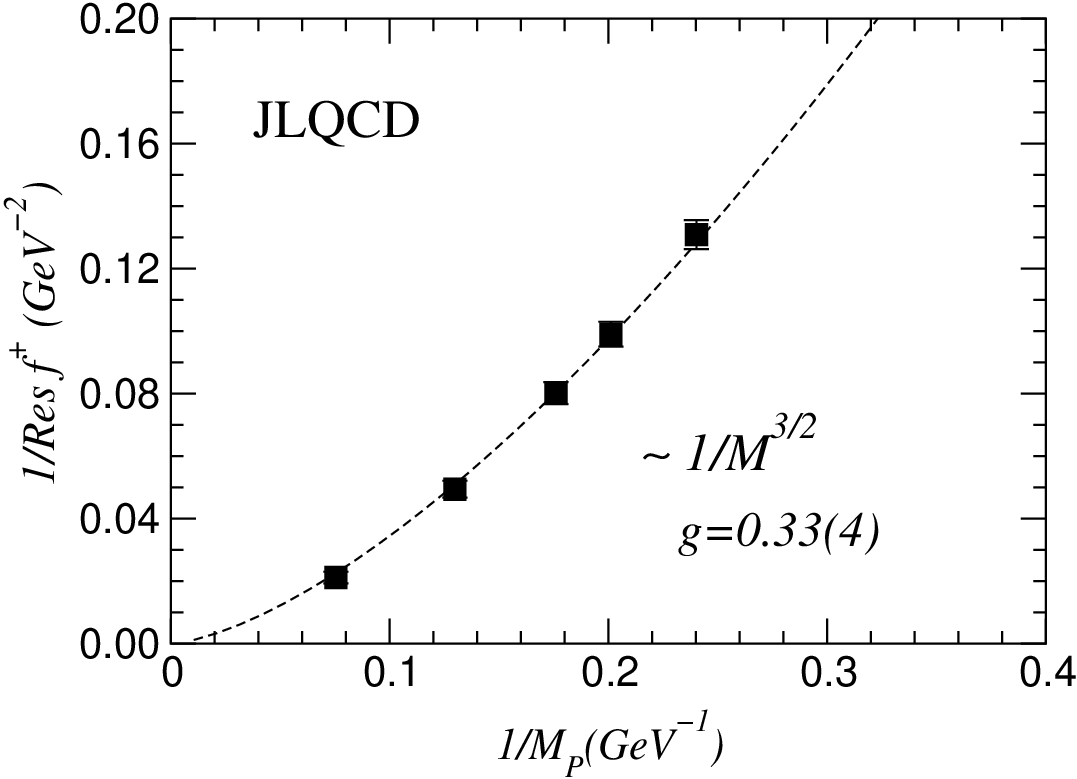}
    \vspace*{-10mm}
    \caption{Heavy quark scaling of the residue of the pole
      dominance model.
      }
%    \vspace*{-5mm}
    \label{fig:f+_scaling}
  \end{center}
\end{figure}

Many previous lattice calculations supported or assumed the
shape of the pole model
\cite{ELC_B2pi_93,APE_B2pi_94,UKQCD_B2pi_95,Hiroshima_B2pi_97}.
At this conference JLQCD has further tested the heavy quark
scaling of the pole model using the NRQCD action
\cite{JLQCD_B2pi_lat99}. 
They fitted observed the measured form factor with
(\ref{eq:pole_dominance}) and extracted the numerator (pole
residue $\mathrm{Res} f^+$). 
The heavy quark scaling law predicts $g_{B^*B\pi}\sim M$ and 
$f_{B^*}\sim M^{-1/2}$, so that the pole residue should
behave as $M^{3/2}$.\footnote{
  This scaling law is compatible with the scaling
  predicted for $f^+(q^2_{max})$ itself ($\sim M^{1/2}$), if 
  the daughter pion mass is kept finite.
  The denominator of (\ref{eq:pole_dominance}) behaves as
  $(m_{B^*}^2-m_B^2)+2m_B m_{\pi} \sim 2 M m_{\pi}$.
  }
They confirmed this behavior as shown in Figure
\ref{fig:f+_scaling}, and obtained a very preliminary
result $g$=0.33(4) for the $B^*B\pi$ coupling, where $g$ is
defined through $g_{B^*B\pi}=(2m_B/f_{\pi})g$. 

A direct lattice calculation of the $B^*B\pi$ coupling has
recently been made by UKQCD collaboration
\cite{UKQCD_B*Bpi_98}.
They use the reduction formula to relate the matrix element
$\langle B(p)\pi(k)|B^*(p')\rangle$ to a `semi-leptonic
transition' amplitude 
$\langle B(p)|A_{\mu}|B^*(p+k)\rangle$, where 
$A_{\mu}$ is a light-light axial current.
Then they use the stochastic propagator to evaluate the
latter matrix element.
Their result $g$=0.42(4)(8) is consistent with a
phenomenological determination through $D^*\rightarrow D\pi$ 
\cite{Stewart_98} and other phenomenological model
calculations \cite{Casalbuoni_et_al_97}.

In the other limit $q^2=0$, where the pion recoil momentum is
large, the light cone scaling law $f^+(0)\sim M^{-3/2}$
holds, which is not compatible with the pole dominance
model that predicts $\sim M^{-1/2}$. 
Becirevic and Kaidalov proposed a model to parametrize
$f^+(q^2)$, introducing a term representing the effects of
higher excited state contributions
\cite{Becirevic_Kaidalov_99}. 
By choosing a parameter of the new term it is possible to
produce an additional suppression of $\sim M^{-1}$, and the 
model becomes consistent with the light cone scaling law.
Using this and other models, such as the pole-dipole model, one
may extrapolate the lattice data to $q^2=0$.
APE \cite{APE_B2pi_lat99} and UKQCD \cite{UKQCD_B2pi_lat99}
presented $f^+(0)=f^0(0)$, which are consistent with each
other and with previous calculations
\cite{Flynn_Sachrajda_97}. 

As I discussed before, the comparison with experiment and
extraction of $|V_{ub}|$ is possible with partial decay
rate without introducing model dependence.
Fermilab group proposed to compare the differential decay
rate $d\Gamma/d|p_{\pi}|$ in the region
400 MeV $\leq |p_{\pi}| \leq$ 850 MeV, where systematic
error is minimized \cite{Fermilab_B2pi_lat98}.
An update was reported at this conference
\cite{Fermilab_B2pi_lat99}, where they found good scaling
between $\beta$=5.7 and 5.9.

\section{$B-\bar{B}$ mixing}
\label{sec:B-B_mixing}

The mass difference of two neutral $B_d$ mesons is measured
quite precisely $\Delta M_d$ = 0.481$\pm$0.017 ps$^{-1}$,
and a bound is known for $B_s$ meson 
$\Delta M_s >$ 14.3 ps$^{-1}$ \cite{LEP_B_Oscillation}. 
To extract the CKM matrix elements $|V_{td}|$ and
$|V_{ts}/V_{td}|$, hadronic parameters $f_B^2 B_B$ and
$f_{B_s}^2 B_{B_s}/f_B^2 B_B$ must be obtained
theoretically, for which the lattice calculation has been
proven to be the best tool. 
Here I describe updates on the calculation of these
quantities. 

\subsection{$f_B$}
\label{sec:f_B}

\begin{figure}[tb]
  \begin{center}
%    \vspace*{-0mm}
    \epsfxsize=74mm \epsfbox{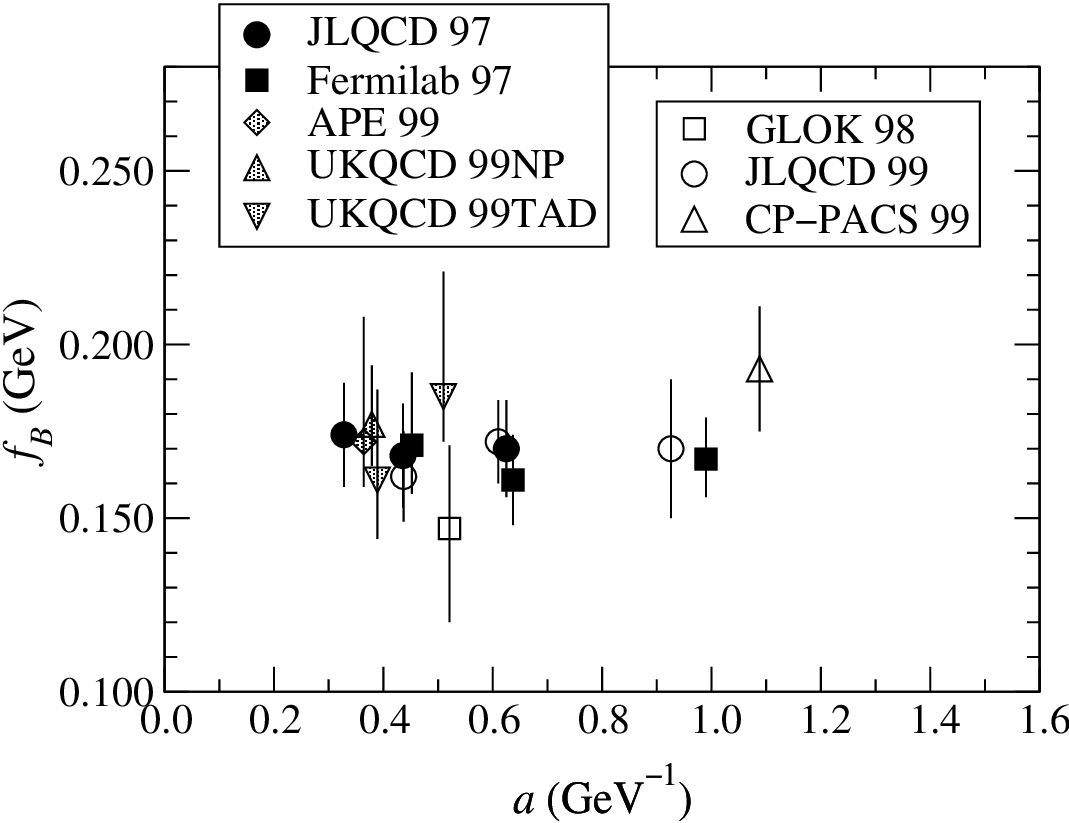}
    \vspace*{-10mm}
    \caption{Recent quenched lattice calculations of $f_B$
      using $O(a)$-improved actions. 
      Results with the Fermilab formalism of heavy quarks
      \cite{El-Khadra_Kronfeld_Mackenzie_97} are
      given by filled symbols:
      JLQCD \cite{JLQCD_fB_97},
      Fermilab \cite{Fermilab_fB_97}.
      Shaded symbols represent calculation involving an
      extrapolation in heavy quark mass:
      APE \cite{APE_fB_98},
      UKQCD \cite{UKQCD_B2pi_lat99,UKQCD_BB_rel_HF8}.
      NRQCD results are given by open symbols:
      GLOK \cite{GLOK_fB_98},
      JLQCD \cite{JLQCD_fB_99},
      CP-PACS \cite{CP-PACD_fB_NRQCD_lat99}.
      }
%    \vspace*{-5mm}
    \label{fig:fB_quenched}
  \end{center}
\end{figure}

In the quenched approximation, the lattice results have
stabilized very well \cite{Draper_lat98}; many groups agree
with each other using different formulation/approach for
heavy quark and different lattice spacings as shown in
Figure \ref{fig:fB_quenched}, where recent quenched results
with $O(a)$-improved actions are plotted as a function of
$a$.
The MILC collaboration has also done an extensive study with
unimproved action, and their result extrapolated to the
continuum limit is consistent with the above improved action 
results \cite{MILC_fB_98}.
The dominant source of systematic errors (apart from
quenching) depends on method and parameters used, but their
typical size is about 10\%.

Then a natural question is what happens with unquenching.
Before this conference, two groups (MILC \cite{MILC_fB_98},
Collins \textit{et al.} \cite{Collins_et_al_fB_99}) reported
dynamical ($N_F=2$) simulations, and both suggested increase 
of $f_B$ with unquenching.\footnote{
  Other theoretical studies, such as Quenched Chiral 
  Perturbation Theory \cite{QChPT}, quark model, `bermions'
  \cite{bermion}, suggested the same conclusion.}

At this conference MILC updated their calculation with three 
new dynamical lattices at $\beta$=5.6 (three different sea
quark masses), which is consistent with their previous
dynamical result \cite{MILC_fB_lat99}. 
A problem in their result is a large $a$ dependence seen
in the quenched data due to the use of unimproved Wilson
quark.
A linear extrapolation to the continuum limit gives a
substantially lower value compared to the data at finite
$a$.
On the other hand, their dynamical results do not show a
similar $a$ dependence and the continuum limit remains
high.
For this reason, although their result suggests 
$f_B^{N_F=2} > f_B^{N_F=0}$, the conclusion is not solid
enough. 
Therefore, they started a new calculation using the
fatlink clover action for heavy quark, with which scaling
behavior is expected to be improved.
\footnote{Necessary one-loop calculation has been performed
  by Bernard and DeGrand
  \cite{Bernard_DeGrand_lat99}.
  }
A preliminary result favors lower value of $f_B$,
albeit with large statistical error.

The CP-PACS collaboration presented two new calculations of 
$f_B$ on their dynamical lattices ($N_F$=2) generated with
an RG improved gauge action
\cite{Burkhalter_CP-PACS_lat98}:

Ali Khan discussed a NRQCD calculation at $\beta$=1.95
($1/a\sim$ 1 GeV) with two sea quark masses
\cite{CP-PACS_fB_NRQCD_lat99}. 
A correction of order $\alpha_s/M$ of the heavy-light
current induced by the operator mixing
\cite{Morningstar_Shigemitsu_98} is included.
\footnote{Necessary one-loop calculation with the RG
  improved gauge action has been performed by Ishikawa 
  \cite{Ishikawa_et_al_lat99}.
  }
They compared the dynamical result with their quenched
result obtained at a similar lattice spacing, and found clear
increase with unquenching ($\sim$ 15--20\%), while no
difference was found between two sea quark masses.

Shanahan presented another systematic study of unquenching
at three $\beta$ values ($1/a$=0.7$\sim$1.7 GeV) with four
sea quark masses (for each $\beta$)
\cite{CP-PACS_fB_relativistic_lat99}. 
They used the $O(a)$-improved relativistic action for
both heavy and light quarks and applied the Fermilab
reinterpretation for heavy quark
\cite{El-Khadra_Kronfeld_Mackenzie_97}. 
The chiral extrapolation was performed with
$m_{sea}=m_{valence}$, so the real unquenching was
made for the first time.

\begin{figure}[tb]
  \begin{center}
%    \vspace*{-0mm}
    \epsfxsize=74mm \epsfbox{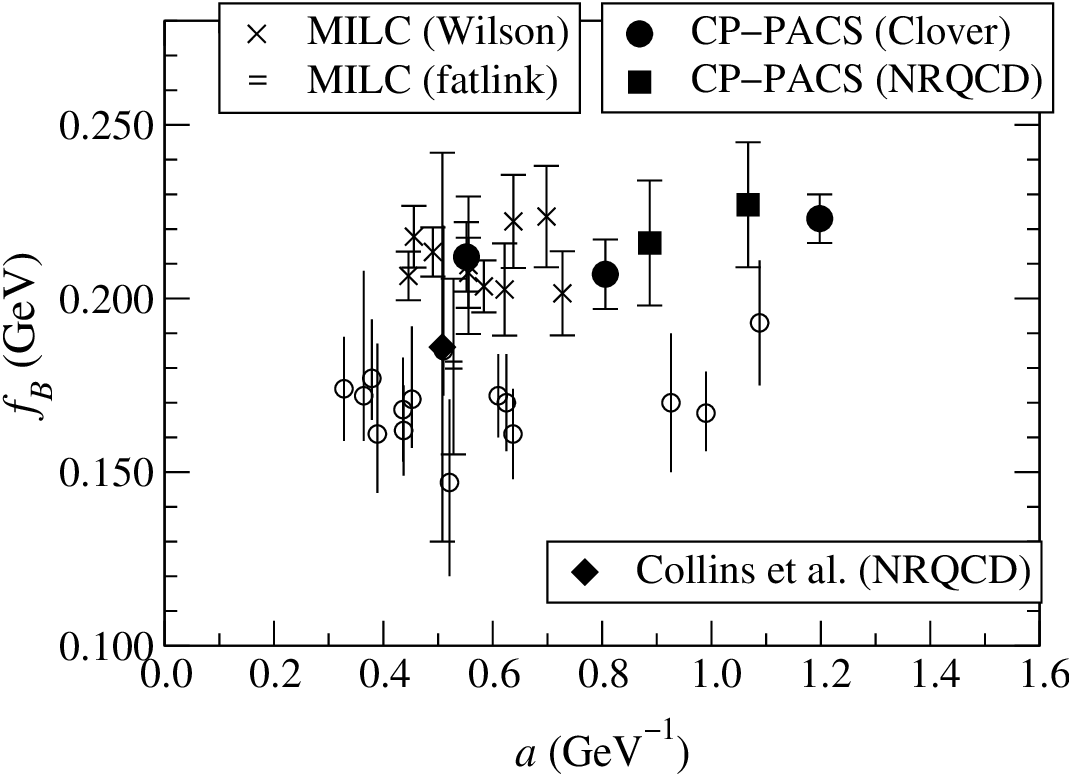}
    \vspace*{-10mm}
    \caption{Dynamical lattice calculations of $f_B$.
      Results are from 
      MILC \cite{MILC_fB_98,MILC_fB_lat99},
      Collins \textit{et al.} \cite{Collins_et_al_fB_99} and 
      CP-PACS
      \cite{CP-PACS_fB_NRQCD_lat99,CP-PACS_fB_relativistic_lat99}. 
      Quenched results as shown in Figure
      \ref{fig:fB_quenched} are also plotted with small open 
      symbols.
      }
%    \vspace*{-5mm}
    \label{fig:fB_dynamical}
  \end{center}
\end{figure}

All dynamical results for $f_B$ are plotted in Figure
\ref{fig:fB_dynamical} together with the recent quenched
data.
We observe clear upward shift of $f_B$ with the inclusion of 
dynamical quarks.
Although it is a difficult task to combine the results from
different groups, we can crudely say that all available data
is consistent with the following estimates: 
\begin{center}
  \begin{tabular}{lll}
    & $N_F$=2 & $N_F$=0 \\
    $f_B$     (MeV) & 210 $\pm$ 30 & 170 $\pm$ 20 \\
    $f_{B_s}$ (MeV) & 245 $\pm$ 30 & 195 $\pm$ 20 \\
    $f_{B_s}/f_B$   & 1.16 $\pm$ 4 & 1.15 $\pm$ 4
  \end{tabular}
\end{center}
where I also list the results for $f_{B_s}$ and
$f_{B_s}/f_B$. 
I do not attempt to extrapolate these results to the physical
$N_F=3$ limit.
To do so, it seems necessary to understand the systematic
errors coming from the use of different actions and lattice
spacings. 
The sea quark mass dependence should also be clarified.

\subsection{$B_B$}
\label{sec:B_B}

In contrast to the achievement for $f_B$, the lattice
calculation of $B_B$ is still premature.

In the static approximation, the $O(a)$-improved results by
Gim\'enez and Martinelli \cite{Gimenez_Martinelli_96} and by
UKQCD \cite{UKQCD_BB_96} have been reanalyzed in a recent
paper by Gim\'enez and Reyes \cite{Gimenez_Reyes_98} using
corrected one-loop matching coefficient, and a
disagreement, which existed between the static-clover and
static-Wilson results\cite{Kentucky_96}, has been greatly
reduced.\footnote{
  UKQCD has also presented a corrected analysis
  \cite{UKQCD_lifetime_98}. But I do not use their number in 
  this summary, since the tadpole improvement is not
  performed in their analysis and there seems large
  systematic error remaining.}

The Hiroshima group performed a calculation using the NRQCD
action \cite{Hiroshima_BB_99,Hiroshima_BB_lat99}, and found
significant decrease of $B_B(m_b)$ as one includes $1/M$
corrections. 
The matching, however, was done with the coefficient in the
static limit, and thus large $O(\alpha/(aM))$ systematic
error is expected for the slope in $1/M$.

In the calculation with relativistic actions, UKQCD
presented the first calculation with the $O(a)$-improved
action at $\beta$=6.0 and 6.2 at Lattice 98
\cite{UKQCD_BB_rel_lat98}, which has recently been updated
\cite{UKQCD_BB_rel_HF8}. 
To obtain the result at the $B$ meson mass, an extrapolation from
charm mass regime is necessary, and they found a clear
negative slope in $1/M$. 

At this conference, APE group \cite{APE_BB_rel_lat99} has
presented the first result obtained using nonperturbative
renormalization \cite{Martinelli_et_al_95}.
They found a similar dependence of $B_B$ on $1/M$, but their
final numerical results are not yet available at 
the time I wrote this contribution.

\begin{figure}[tb]
  \begin{center}
%    \vspace*{-0mm}
    \epsfxsize=74mm \epsfbox{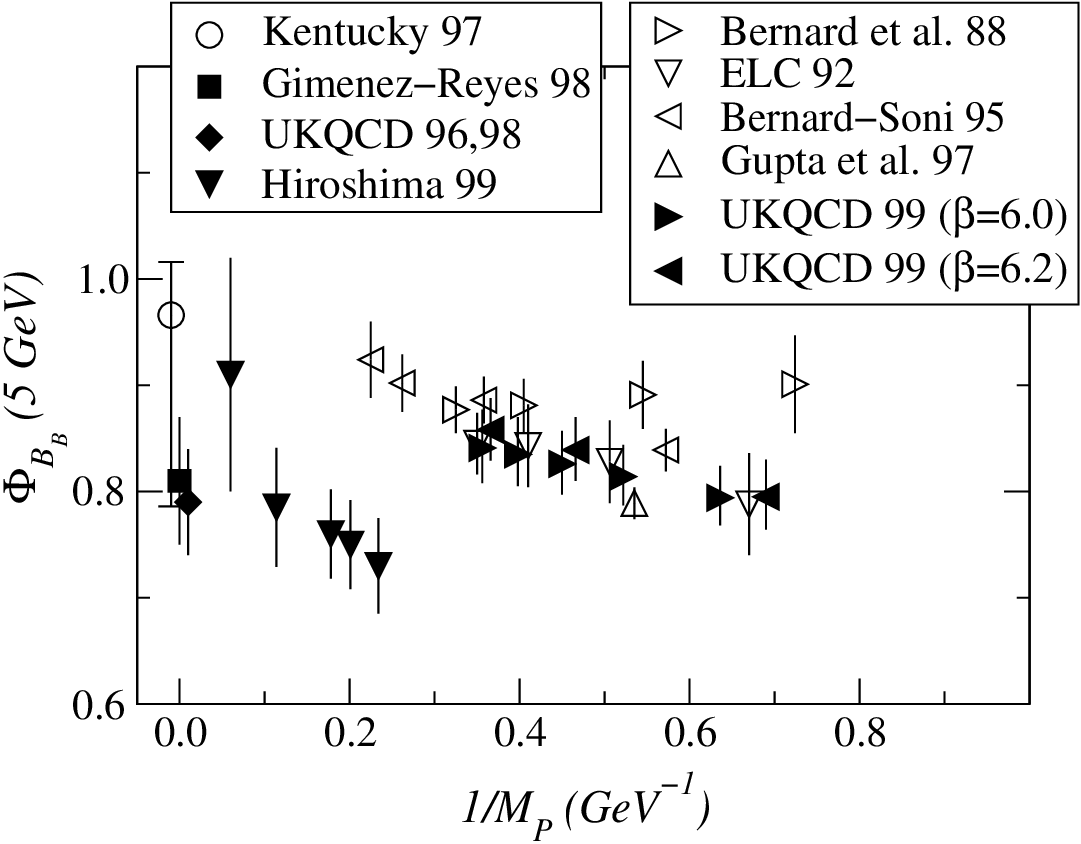}
    \vspace*{-10mm}
    \caption{$1/M$ dependence of $\Phi_{B_B}(5 GeV)$.
      The static and NRQCD data are from
      Kentucky \cite{Kentucky_96},
      Gim\'enez-Reyes \cite{Gimenez_Reyes_98} (Reanalysis of 
      \cite{Gimenez_Martinelli_96}),
      UKQCD \cite{UKQCD_BB_96} (Reanalyzed by
      \cite{Gimenez_Reyes_98}), and
      Hiroshima \cite{Hiroshima_BB_99}.
      The relativistic calculations are
      Bernard \textit{et al.} \cite{Bernard_et_al_88},
      ELC \cite{ELC_92},
      Bernard-Soni \cite{Soni_lat95},
      Gupta \textit{et al.} \cite{Gupta_et_al_97}, and 
      UKQCD \cite{UKQCD_BB_rel_lat98}.
      Open symbols are obtained with Wilson quark for heavy
      and/or light quarks, and filled ones are
      $O(a)$-improved. 
      }
%    \vspace*{-5mm}
    \label{fig:B_B}
  \end{center}
\end{figure}

Figure \ref{fig:B_B} presents a compilation of lattice data
for 
$\Phi_{B_B}(\mu_b)\equiv 
(\alpha_s(M_P)/\alpha_s(M_B))^{2/\beta_0} B_B(\mu_b)$ 
with 
$\mu_b$=5 GeV as a function of $1/M_P$. 
The renormalization factor is introduced to cancel the 
$\ln(M/\mu_b)$ dependence appearing in the matching factor 
\cite{UKQCD_BB_rel_HF8}. 
It is encouraging that all relativistic results including 
the early works \cite{Bernard_et_al_88,ELC_92} show a  
reasonable agreement with each other, and that the recent
UKQCD data show a nice scaling between $\beta$=6.0 and 6.2. 
The extrapolation to the static limit ($\sim$ 0.92), however, 
seems considerably higher than the $O(a)$-improved results
in that limit. 
It suggests that there is unknown sources of systematic
error in either or both of static (NRQCD) and relativistic
calculations. 
Higher order perturbative corrections (for both) and 
$O((aM)^2)$ uncertainty in the relativistic calculations
are their potential candidates.
For this reason, my summary of the current available data
includes a large systematic uncertainty: 
$B_B(m_b)$ = 0.80(15).

\begin{figure}[tb]
  \begin{center}
%    \vspace*{-0mm}
    \epsfxsize=74mm \epsfbox{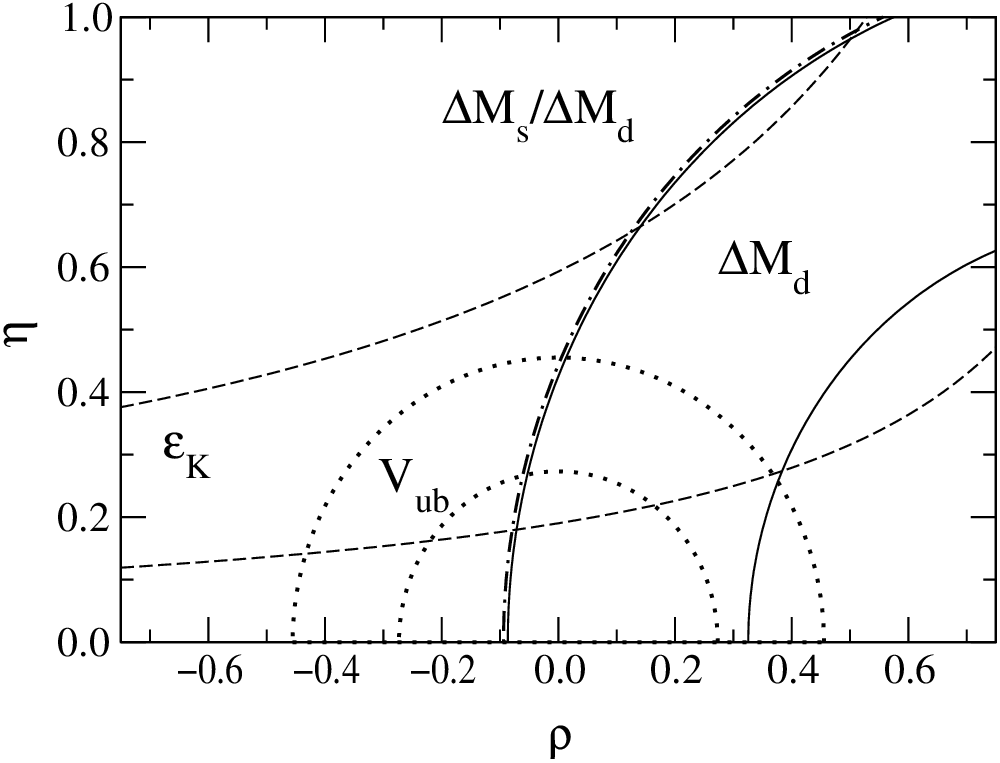}
    \vspace*{-10mm}
    \caption{Constraint on the CKM matrix elements.}
%    \vspace*{-5mm}
    \label{fig:CKM}
  \end{center}
\end{figure}

The allowed region on the $(\rho,\eta)$ plane of the CKM matrix
is shown in Figure \ref{fig:CKM}. 
The two flavour result $f_B$ = 210 $\pm$ 30 MeV and the 
conservative estimate of $B_B$ are used to draw the
constraint from $\Delta M_d$.
Due to the upward shift of $f_B$ from the previous quenched 
results, the allowed region favors $\rho >$ 0. 

\subsection{$B_s$ width difference}
\label{sec:B_s_width_difference}
 
The width difference in the $B_s-\bar{B}_s$ mixing is given
as 
\begin{eqnarray}
  \label{eq:width_difference}
  \lefteqn{\Delta\Gamma_s \propto} \nonumber \\
  & &
  Im \frac{1}{2M_{B_s}}
  \langle\bar{B}_s| i\int d^4x T
  \mathcal{H}^{\mathrm{eff}}(x)
  \mathcal{H}^{\mathrm{eff}}(0) 
  |B_s\rangle,
\end{eqnarray}
where $\mathcal{H}^{\mathrm{eff}}$ represents the 
$\Delta B$=1 effective Hamiltonian
\cite{Beneke_Buchalla_Dunietz_96}. 
Only the final states into which both of $B_s$ and
$\bar{B}_s$ can decay contribute.
The $1/m_b$ expansion induces two four-quark $\Delta B$=2
operators, whose matrix elements with $B_s$ and $\bar{B}_s$
states are $B_B$ and $B_S$, where
$B_S$ is defined through
\begin{eqnarray}
  \label{eq:B_S}
  \lefteqn{\langle\bar{B}_s|\mathcal{O}_S(\mu)|B_s\rangle =}
  \nonumber \\
  & &
  -\frac{5}{3} f_{B_s}^2 M_{B_s}^2 
  \frac{M_{B_s}^2}{(\overline{m}_b+\overline{m}_s)^2}
  B_S(\mu),
\end{eqnarray}
and $\mathcal{O}_S=\bar{b}(1-\gamma_5)s\bar{b}(1-\gamma_5)s$.

At this conference, the Hiroshima group presented a calculation
of $B_S$ using the NRQCD action \cite{Hiroshima_BB_lat99}.
Their calculation method is the same as that of $B_B$ and
they obtain $B_S(m_b)$=1.19(2)(20).
Using a next-to-leading order formula of 
Beneke \textit{et al.} \cite{Beneke_et_al_98}, the width
difference is obtained as
$(\Delta\Gamma/\Gamma)_{B_s}$ =  0.16(3)(4),
where errors are from $f_{B_s}$ and $B_S$ respectively.
The two-flavour result for $f_{B_s}$ discussed in Section
\ref{sec:f_B} is used.
The latest experimental bound from DELPHI is
$(\Delta\Gamma/\Gamma)_{B_s} <$ 0.42 \cite{DELPHI_BS_99}.

\subsection{Lifetime ratios}
\label{sec:Lifeteime_ratios}

The ratios of lifetime of $b$ hadrons, such as
$\tau(B^-)/\tau(B^0)$ and $\tau(\Lambda_b)/\tau(B^0)$,
provide an important test of the theoretical method to
calculate the inclusive hadronic decay rates
\cite{Neubert_Sachrajda_96}.
In the $1/m_b$ expansion, the leading contribution to the
decay rate comes from a diagram in which the $b$-quark decay
proceeds without touching the spectator quark, so that it
does not contribute to the lifetime ratios.
The $O(1/m_b^2)$ correction to the ratios is also small for
the same reason, and the first correction involving the
spectator quark effect is of $O(1/m_b^3)$, which is
parametrized by the `$B$ parameters' of $\Delta B$=0
four-quark operators.
UKQCD computed these matrix elements for the first time and
obtained 
$\tau(B^-)/\tau(B^0)$ = 1.03(2)(3) \cite{UKQCD_lifetime_98}, 
which is consistent with the recent experimental result 
1.07(2) \cite{LEP_B_Lifetime}. 

It is a known problem that the lifetime of $\Lambda_b$
is surprisingly shorter than that of $B$ mesons
$\tau(\Lambda_b)/\tau(B^0)$ = 0.79(5) \cite{LEP_B_Lifetime}.
It is, therefore, interesting to see whether it is explained 
with the theoretical calculation, in which the similar
matrix elements of four quark operators for $\Lambda_b$ are
required. 
The UKQCD group has studied these matrix elements at
$\beta$=5.7 with 20 gauge configurations
\cite{UKQCD_lifetime_99}, and found that the spectator 
effect is large $\sim -$6\%.
Although their result $\tau(\Lambda_b)/\tau(B^0)$ =
0.91(1)$\sim$0.93(1), depending on the light quark mass, is
much higher than the experimental value, higher statistics
calculations at higher $\beta$ values seem necessary to draw 
a definite conclusion.

\section{Conclusions}
\label{sec:Conclusions}

Lattice calculations provide model independent 
predictions for many important $B$ decay matrix elements. 
Progress made for the zero recoil 
$B\rightarrow D^{(*)}l\nu$ from factors is essential for 
precise determination of $|V_{cb}|$, and the shape of the
form factors is also being studied with improved techniques 
(NP improved action, NRQCD, etc.).
More study is necessary to achieve a complete understanding of 
the $B\rightarrow \pi l\nu$ form factors: the violation of
soft pion theorem and the shape of $f^+(q^2)$.
The determination of $|V_{ub}|$ with 10\% precision will
become possible, once we understand these questions.

The dynamical quark simulation has become practical by
several groups, and its effect on $f_B$ has been identified. 
Further systematic study like that of MILC and CP-PACS is necessary
to understand systematic errors and eventually to obtain
physical prediction at $N_F=3$.
An unquenched study of other quantities should also be
important. 

Several new applications have also been studied, such as 
the width difference of $B_s$ meson, and the lifetime ratios
of $b$ hadrons. 
Those will become useful theoretical calculations, provided 
that statistical and systematic errors are improved.

\section*{Acknowledgements}

I thank
A.~Ali~Khan, 
S.~Aoki,
D.~Becirevic,
C.~Bernard,
S.~Collins,
C.~DeTar,
G.~Douglas, 
V.~Gim\'enez,
L.~Giusti,
J.~Hein,
K-I.~Ishikawa, 
A.~Kronfeld, 
L.~Lellouch,
C.-J.D.~Lin,
C.~Maynard,
T.~Onogi, 
S.~Ryan,
H.~Shanahan, 
J.~Simone, and
N.~Yamada
for communications and useful discussions.
I also thank M.~Okawa and A.~Ukawa for comments on the
manuscript. 
S.H. is supported in part by the Ministry of Education under
Grant No. 11740162.

\end{document}